\documentclass[aps,prl,twocolumn,epsfig,floats,superscriptaddress]{revtex4}
\usepackage{graphicx,epsf}
\def\be{\begin{equation}}
\def\ee{\end{equation}}
\def\bea{\begin{eqnarray}}
\def\eea{\end{eqnarray}}

\begin{document}


\title{Odd parity superconductivity in Weyl semimetals}
\author{Huazhou Wei}
\affiliation{Department of Physics and Astronomy, University of California,
Riverside, CA 92521}
\author{Sung-Po Chao}
\affiliation{Physics Division, National Center for Theoretical Science, Hsinchu, 30013, Taiwan, R.O.C.}
\affiliation{Physics Department, National Tsing Hua University, Hsinchu, 30013, Taiwan, R.O.C.}
\author{Vivek Aji}
\affiliation{Department of Physics and Astronomy, University of California,
Riverside, CA 92521}
\begin{abstract}
Unconventional superconducting states of matter are realized in the presence of strong spin orbit coupling. In particular, non degenerate bands can support odd parity superconductivity with rich topological content. Here we study whether this is the case for Weyl semimetals. These are systems whose low energy sector, in the absence of interactions, is described by linearly dispersing chiral fermions in three dimensions. The energy spectrum has nodes at an even number of points in the Brillouin zone. Consequently both intranodal finite momentum pairing and internodal BCS superconductivity are allowed. For local attractive interaction the finite momentum pairing state with chiral p-wave symmetry is found to be most favorable at  finite chemical potential. The state is an analog of the superfluid $^{3}$He A phase, with cooper pairs having finite center of mass momentum. For chemical potential at the node the state is preempted by a fully gapped charge density wave. For long range attraction the BCS state wins out for all values of the chemical potential. 
\end{abstract}
\maketitle

 Weyl semimetals are a class of materials whose low energy description is in terms of linearly dispersing massless fermions in three dimensions. They occur in systems where two non degenerate bands touch at isolated points in the Brilloiun zone. In general the existence of such Weyl nodes at the chemical point is accidental\cite{herr}, but if they do exist then they are robust against perturbations that do not break translational invariance. The presence of degenerate chiral nodes separated in momentum space requires that either inversion or time reversal symmetry be broken. There are a number of nontrivial consequences such as open Fermi surfaces for surface states, anomalous Hall effects and other transport features\cite{xwan, bb, hosur, abjaji, zyuzin3,pallab}. These phenomena can be traced back to the conservation of chirality at the Weyl nodes. While Weyl fermions have been studied extensively in the context of liquid $^{3}$He, the recent proposals of realizing them in pyrochlore iridates\cite{xwan} and Topological-Normal Insulator (TNI) heterostructure\cite{bb} has renewed interest on the subject.  
 
 Of particular interest is the nature of correlated phases they support. For chemical potential at the nodes, perfect nesting leads to possible particle-hole instabilities. For local repulsion, an excitonic ferromagnetic insulator is stabilized\cite{hwei}, while for local attractive interactions a Charge Density Wave (CDW)\cite{zwang} is realized. Meng and Balents\cite{meng} studied the nature of superconducting state obtained in systems where the superconductivity is externally induced by proximity effect. This is achieved by replacing the normal insulator by a superconductor in the Topological-Normal insulator (TNI) heterostructure. They find a variety of gapless and/or topological superconducting phases which may host Majorana bound states on the surface or vortex cores. Cho et al.\cite{gyc} studied the intrinsic superconducting instabilities of doped Weyl semimetals within a model that has $C_{4h}$ point group symmetry. They find that the even parity fully gapped finite momentum pairing state is energetically favored. The point group symmetry imposed is not necessary for Weyl semimetals. In this work we relax this constraint and explore the possible superconducting phases. 
 
 For local attractive interactions we find the finite momentum pairing to be the ground state, while for long range interaction a gapped BCS state is a competing phase, with details of the interaction favoring one over the other. Crucially, contrary to Cho et al. \cite{gyc} we find that a "spin singlet" has no weight and that only p-wave "spin triplet" phases are allowed. The difference originates from the properties of the model under inversion. In our case the spin at momentum $\vec{k}$ and momentum $-\vec{k}$ are the same as required by inversion symmetry. On the other hand, even for inversion symmetric models, the effective low energy theory can be one where the spins are not parallel at momenta related by inversion\cite{bb}. In the latter case, singlet pairing has finite overlap with the chiral state. For the class of Weyl semimetals studied here, we generically find odd parity superconductivity which are analogs of the $^{3}He$ A phase. They add to a class of spin triplet superconducting phases that display Weyl behavior \cite{liwu, sauti}
  
The approach we take is the same as the one used to explore excitonic phases\cite{hwei}. In this regard the work is complementary to that of Cho et al.\cite{gyc}, who look at mean field decomposition in the spin basis prior to projecting to the low energy sector. We first project to the linearly dispersing chiral basis and then perform the mean field analysis. To highlight the physics, we simplify to the case of two Weyl nodes with density density interactions. There are two types of particle-particle instabilities that can arise in this case i) intra-nodal (occurring at zero momentum) and ii) inter-nodal (occurring at a finite fixed momentum associated with the nesting vector). The former leads to finite momentum pairing (analogous to FFLO\cite{ff,lo}) while the latter is the zero momentum pairing BCS\cite{bcs} state. For local interaction, the most favorable superconducting state is the finite momentum paired odd parity axial phase. A minimum interaction strength is required to nucleate them for chemical potential at the node which is the consequence of the vanishing density of states. We also find that it is energetically less optimal than a fully gapped CDW phase. For finite chemical potential the particle hole nesting is lost, and the axial superconductor is realized. For long range attraction a fully gapped BCS state is stabilized for all values of the chemical potential.

\noindent\underline{\textit {Model}}: Consider a system with two Weyl nodes at $\vec{K}_{0} = K_{0}\hat{x}$ (labeled R) and $-\vec{K}_{0}= -K_{0}\hat{x}$ (labeled L) with chiralities $+1$ and $-1$ respectively. The Hamiltonian is

\begin{equation}
H_{0 \pm} = \pm \hbar v \sum_{\vec{k}}\psi_{\vec{k}\alpha}^{\dagger}\vec{\sigma}_{\alpha\beta}\cdot\left(\vec{k}\mp \vec{K}_{0}\right)\psi_{\vec{k}\beta},
\end{equation}
where $v$ is the Fermi velocity and $\vec{\sigma} = \{\sigma_{x},
\sigma_{y}, \sigma_{z}\}$ is a vector of Pauli matrices. The
dispersion at each node is $\epsilon_{\vec{q}}=\pm \hbar
v\left|\vec{q}\right|$ centered around $\pm \vec{K}_{0}$, with
$\vec{q} =\left(\vec{k}\mp \vec{K}_{0}\right)$. The conduction
(valence) band at the R node has its spin parallel (anti-parallel)
to $\vec{q}$, while the opposite is true at the L node. The general
particle particle interaction, in momentum space, takes the form
\begin{equation}\label{vg}
V=\sum_{\sigma,\sigma'}\sum_{\vec{k}, \vec{k}',\vec{q}}V(\vec{q}) \psi^{\dagger}_{\vec{k'}+\vec{q},\sigma'} \psi_{\vec{k'},\sigma'}\psi^{\dagger}_{\vec{k}-\vec{q},\sigma}\psi_{\vec{k},\sigma}
\end{equation}

\begin{figure}
\includegraphics[width=0.6\columnwidth]{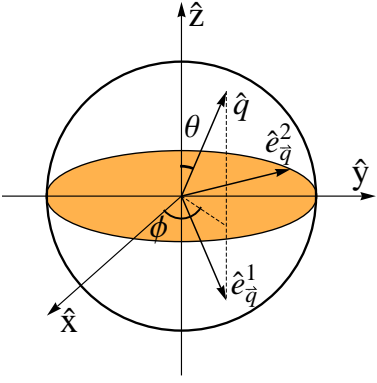}
\caption{The interaction shown in Eq.(\ref{cooper}) is a function of three vectors ($\hat{q}$, $\hat{e}^{1}_{\vec{q}}$ and $\hat{e}^{2}_{\vec{q}}$) that form a right handed coordinate system. Each vector couples to an operator of distinct symmetry in the particle hole channel.}
 \label{vectors}
\end{figure}

\noindent Since the Weyl physics is the low energy description of a more general theory, we enforce an upper cutoff in the momentum integrals (up to an energy $\Lambda$) around the Weyl point.

\noindent\underline{\textit {Particle-Particle instabilities}} :  We rewrite the interaction in the basis of the non-interacting bands. To do so we define a rotation matrix $M^{R,L} (\vec{k})_{n\sigma}$ such that $c^{L,R}_{\vec{k}n} = M^{L,R}(\vec{k})_{n\sigma}\psi^{R,L}_{\vec{k}\sigma}$. Note that the spin degeneracy is lifted and the noninteracting eigenstates are labels by the band index $n=\pm$.  The rotation matrices are unitary and rotate the spin quantization axis of each electron to point along its momentum $\vec{k}$. 

We split the sum over momentum over $\vec{k}$ for each $\psi_{\vec{k},\sigma}$ into two, one with small momenta near the left node and the other with small momenta about the right node. An upper cutoff in energy, $\Lambda$, is imposed as the linear dispersion is a low energy phenomena. Of the $16$ possible terms from Eq.(\ref{vg}) only $6$ terms satisfy momentum conservation for scattering restricted to the states within the cutoff around the node. For every momentum $\vec{q}=q\hat{q}$, where $\hat{q} = \{\hat{q}_{x}, \hat{q}_{y}, \hat{q}_{z}\}$ is the unit vector along $\vec{q}$, we define two orthogonal vectors $\hat{e}_{\vec{q}}^{1} \equiv \hat{\theta}_{\vec{q}} = \{\hat{q}_{x}\hat{q}_{z}/\sqrt{\hat{q}_{x}^{2}+\hat{q}_{y}^{2}},\hat{q}_{y}\hat{q}_{z}/\sqrt{\hat{q}_{x}^{2}+\hat{q}_{y}^{2}}, -\sqrt{\hat{q}_{x}^{2}+\hat{q}_{y}^{2}}\}$ and $\hat{e}_{\vec{q}}^{2}\equiv \hat{\phi}_{\vec{q}} = \{-\hat{q}_{y}/\sqrt{\hat{q}_{x}^{2}+\hat{q}_{y}^{2}}, \hat{q}_{x}/\sqrt{\hat{q}_{x}^{2}+\hat{q}_{y}^{2}},0\}$, such that $\hat{q}$, $\hat{e}_{\vec{q}}^{1} $ and $\hat{e}_{\vec{q}}^{2} $ form a right handed coordinate system (see Fig.\ref{vectors}). The unit sphere is spanned by the vector $\hat{q}$ by two rotations, one about any axis perpendicular to $\hat{e}^{2}_{\vec{q}}$ and the another about $\hat{e}_{\vec{q}}^{2}$. 
Construction above holds for an arbitrary quantization axis $\hat{n}$, with the corresponding polar and azimuthal angle for $\vec{q}$ defined in the coordinate frame  $\{\hat{l}, \hat{m}, \hat{n}\}$. In the rest of the letter we use the $\{\hat{x}, \hat{y}, \hat{z}\}$ coordinate system. The particular choice of the coordinate system breaks spatial rotational invariance.

Specializing to potentials that are even functions of $\vec{k}$, i.e. $V(\vec{k}) = V(-\vec{k})$, the interaction is
\begin{widetext}
 \begin{eqnarray}\label{cooper}
 V_{c} &=& \sum_{\vec{k},\vec{k'},\tau,n}\left[ \right.    {V(\vec{k}-\vec{k}')-V(\vec{k}+\vec{k}'-2\tau \vec{K}_{0})\over 2}(\hat{e}^{1}_{\vec{k}}\cdot \hat{e}^{1}_{\vec{k}'}+\hat{e}^{2}_{\vec{k}}\cdot \hat{e}^{2}_{\vec{k}'})(c_{\vec{k}n}^{\tau \dagger}c_{\vec{-k}n}^{-{\tau} \dagger}c_{-\vec{k}'n}^{-{\tau}}c_{\vec{k}'n}^{\tau})\nonumber \\&+&{V(\vec{k}-\vec{k}')\over 2}(1+\hat{k}\cdot\hat{k}')c_{\vec{k}n}^{\tau \dagger}c_{\vec{-k}n}^{{\tau} \dagger}c_{-\vec{k}'n}^{{\tau}}c_{\vec{k}'n}^{\tau}\left.\right]
 \end{eqnarray}
  \end{widetext}
 $\tau=\pm$ refer to the two (left and right) nodes. We have dropped terms of the form $c_{n\vec{k}}^{\tau_{1}\dagger}c_{-n-\vec{k}}^{\tau_{2}\dagger}c_{-n-\vec{k}'}^{\tau_{2}}c^{\tau_{1}}_{n \vec{k}'}$ and $c_{n\vec{k}}^{\tau_{1}\dagger}c_{n-\vec{k}}^{\tau_{2}\dagger}c_{-n\vec{k}'}^{\tau_{2}}c^{\tau_{1}}_{-n -\vec{k}'}$ which lead to pairing of states which are not degenerate in the noninteracting limit. For attractive interaction we get the very rich structure, with a number of possible superconducting phases. The first two terms in eqn.\ref{cooper} lead to inter nodal pairing, which give the zero momentum BCS\cite{bcs} state, while last term yields finite momentum pairing states (FFLO\cite{ff,lo}). In the rest of the paper we analyze the instabilities within mean field.

\noindent\underline{\textit{Local interactions }:} For local interactions the BCS channel vanishes. To understand why, note that the interaction is one where we destroy particles at $\vec{k}$ and $-\vec{k}$ and create them at $\vec{k}'$ and $-\vec{k}'$. Thus there are two possibilities: put the first particle at $\vec{k}'$ and the second at $-\vec{k}'$ or vice verse. These are inequivalent processes, as evident from the different momentum transfer involved, among indistinguishable particles. The exchange produces a relative minus sign. For local interaction the weight of both the processes are identical leading to an exact cancellation. Note that this result is very different from that of the model with lower symmetry such as the $C_{4h}$ symmetric model studied by Cho et al.\cite{gyc}, where the BCS channel is also unstable for local attraction. The reason for the difference arises from the fact that in our model the spins at $\vec{k}$ and $-\vec{k}$ are parallel for all $\vec{k}$. Thus the two process only pick up a relative sign independent of the spin orientation. For the $C_{4h}$ symmetric models, whether the particle at $\vec{k}$ ends up at $\vec{k}'$ or $-\vec{k}'$ also determines a relative factor that accounts for the different spin orientation of the two final states. This mitigates the cancellation for all momenta. Nevertheless, in both cases the finite momentum pairing wins out. Tipping the system to favor the BCS state requires fine tuning.

In the finite momentum pairing channel, there are two equally attractive channels corresponding to order parameters of the form $\Delta_{s} = \left<\sum_{\vec{k}}c_{-\vec{k}n}^{{\tau}}c_{\vec{k}n}^{\tau}\right>$ and $\vec{\Delta}_{p} = \left<\sum_{\vec{k}}\hat{k}c_{-\vec{k}n}^{{\tau}}c_{\vec{k}n}^{\tau}\right>$. The former is the even parity (s-wave) while the latter is odd parity (p-wave) superconductor. Note that the anti commutation of fermonic operators implies that $\Delta_{s} =0$. This is expected as nondegenrate states cannot pair in the singlet channel and only odd orbital pairing survives. For local attractive interaction, $V(\vec{k}) = g/\Omega$ where $g$ is a constant and $\Omega$ is the volume of the system. The gap equation for the p-wave channel at zero temperature is 
\begin{eqnarray}\label{mf1}
1 &=& {g\over 2}\sum_{k}{\left|\hat{\Delta}_{p}\cdot \hat{k}\right|^{2}\over{\sqrt{(\hbar v k)^{2}+\left|\vec{\Delta}_{p}\cdot\hat{k}\right|^{2}}}}
\end{eqnarray}
In general the complex order parameter takes the form $\vec{d}_{1} + i\vec{d}_{2}$ and extremization yields two possible structure: (i) $\vec{d}_{1}\cdot\vec{d}_{2} = 0$ , $|\vec{d}_{1}|=|\vec{d}_{2}|$ and (ii)  $\vec{d}_{1} || \vec{d}_{2}$, $\vec{d}_{1}+i\vec{d}_{2} = \vec{d}e^{i\phi}$ where $\vec{d}$ is a real vector\cite{volovik}. Minimization of the gap equation for the two cases yields the axial vacuum (case (i)) as the ground state and thus a chiral superconductor is stabilized. This states has nodes in the gap, with linearly dispersing massless charged excitations, in complete analogy with the A phase of liquid $^{3}$He. Equation (\ref{mf1}) is identical to the gap equation obtained for the excitonic phases for repulsive interaction \cite{hwei}. Reading off the results we note that a minimum interaction strength of $g_{c} = 3(\hbar v)^{3}/2\pi\Gamma^{2}$ is required for the state to be realized for chemical potential at the node. Here $\Gamma <\Lambda$ is the cutoff in energy of the attractive interaction. At mean field level, the CDW instability is also possible for attractive interaction \cite{hwei, zwang}. The critical coupling is smaller as compared to the superconducting state and opens a full gap (i.e. no nodes). Thus the nodal finite momentum superconducting state is always disfavored as compared to the CDW. 

At finite chemical potential the particle hole nesting between the nodes is lost and only the superconducting state is realized. Moreover, for finite chemical potential, $\mu$, the state is precipitated for infinitesimal interaction strength. For $\mu-\Gamma>0$ and $\mu+\Gamma<\Lambda$, the attractive interaction is operative only for the positive energy sector of the theory with linear dispersion. For this case the transition temperature is $2 K_{B}T_{c} \approx \Gamma \exp [ -3/g\nu (\mu)]$ where $\nu (\mu) = \mu^{2}/2\pi^{2}(\hbar v)^{3}$ is the density of states at the chemical potential.

\noindent\underline{\textit{Long range interactions}:}  For local interaction $V(\vec{k}-\vec{k}')=V(\vec{k}+\vec{k}'-2\tau \vec{K}_{0})$ and no inter nodal pairing is allowed. For long range interaction, the cancellation does not occur and a BCS state can precipitate. This state competes with the p-wave intra-nodal pairing state. Which of the two wins depends on the details of the interaction. To identify the possible phases, we assume an attractive interaction of the form 

\begin{equation}\label{longint}
V(\vec{k})= \left\{ \begin{array}{rl}
 -g &\mbox {if $|\vec{k}|<|\vec{K}|<|\vec{K}_{0}|$} \\
  0 &\mbox{ otherwise}
       \end{array} \right.
\end{equation}
for some fixed $\vec{K}$. Thus the attraction has a range of order $1/|\vec{K}|$ smaller the $1/|\vec{K}_{0}|$. While this simplifies the algebra, the symmetry arguments below hold in general. Let us now consider the two attractive channels: (1) $\vec{\Delta}_{1} = \left<\sum_{\vec{k}}\hat{e}^{1}_{\hat{k}}c_{-\vec{k}n}^{{\tau}}c_{\vec{k}n}^{\tau}\right>$ and (2) $\vec{\Delta}_{2} = \left<\sum_{\vec{k}}\hat{e}^{2}_{\hat{k}}c_{-\vec{k}n}^{{\tau}}c_{\vec{k}n}^{\tau}\right>$. Since $\hat{e}_{\vec{k}}^{1}$ is even under inversion , i.e. $\hat{e}_{\vec{k}}^{1}$ = $\hat{e}_{-\vec{k}}^{1}$, $\vec{\Delta}_{1}=0$. This is analogous to the even orbital  parity channel vanishing in the intranodal case.

There are two possible superconducting states for $\vec{\Delta}_{2}$: (1) $\vec{\Delta}_{2} = \left<\sum_{\vec{k}}\hat{e}^{2}_{\hat{k}}c_{-\vec{k}n}^{{\tau}}c_{\vec{k}n}^{\tau}\right>$ $=\Delta_{2p}\hat{x} $ and  $\vec{\Delta}_{2} = \left<\sum_{\vec{k}}\hat{e}^{2}_{\hat{k}}c_{-\vec{k}n}^{{\tau}}c_{\vec{k}n}^{\tau}\right>$ $=\Delta_{2c}(\hat{x}+i\hat{y})/\sqrt{2}$. The $p$ and $c$ label refer to polar and chiral respectively. The structure of the order parameters is dictated by symmetry. Once the spatial rotational symmetry is broken by a choice for the quantization axis, the vector $\hat{e}_{2}$ lies in the plane perpendicular to it. As the vector $\hat{k}$ sweeps out the unit sphere, $\hat{e}_{2}$ spans a unit circle in this plane (see fig.\ref{vectors}). Thus the order parameter in this case is either a polar vector in the plane (chosen to be $\hat{x}$ for illustrative purposes) or chiral. Within mean field, the spectrum for the quasiparticles for the two cases are $E_{2p} = \sqrt{(\hbar v k)^{2}+ |\Delta_{2p}|^{2}\cos^{2}\phi}$ and $E_{2c} = \sqrt{(\hbar v k)^{2}+ |\Delta_{2c}|^{2}}$, where $\phi$ is the azimuthal angle in the $\{\hat{x}, \hat{y},\hat{z}\}$ coordinate system. The  polar state has line nodes while the chiral state is gapped. On minimization of the free energy, the latter is energetically favored. It is also more favorable as compared to the finite momentum pairing state.

For chemical potential at the node a minimum coupling strength of $g_{c} = (\hbar v)^{3}/2\pi\Gamma^{2}$ is needed to nucleate this state. Here $\Gamma = \hbar v |\vec{K}|$ is the energy corresponding to the cut off in momentum in eqn.\ref{longint}. Since the intranodal pairing depends only on the small wavelength part of the interaction, the instability criterion is the same for the interaction in eqn.\ref{longint} as the short range interaction. The critical coupling is three times larger in the latter case, so that for long range interactions the chiral BCS state is the preferred ground state. 

For finite chemical potential, the transition temperature for the chiral BCS state is $2 K_{B}T_{c} \approx \Gamma \exp [ -1/2 g\nu (\mu)]$. Thus the transition temperature is lower than that of the finite pairing state given by $2 K_{B}T_{c} \approx \Gamma \exp [ -3/ g\nu (\mu)]$. The difference arises from the angular dependence of the gap in the finite momentum state which has nodes at the poles. 

\noindent\underline{\textit{Topological excitations }:} For short range interactions the lowest energy state is the finite momentum pairing in odd parity channel. Such a state has nodes at the north and south pole of the spherical fermi surface. In complete analogy with the corresponding states for spinless version of the equal spin pairing states in $^{3}$He\cite{volovik}, they support relativistic massless fermonic excitations. The existence  of these nodal points leads to surface states at zero energy. As discussed in refs.[\onlinecite{gyc,df}] the vortex  of finite momentum pairing state is made up of two half quantum vortices, where the phase only winds around one of the Weyl nodes but not the other. The fact that Fermi surface encloses a Berry phase of $\pi$, implies that each half vortex hosts a Majorana mode at its core. In general the hybridization between the two will gap them out as they are not protected by any symmetry. For long range interaction, the odd parity BCS state wins out. This state is fully gapped.

\noindent\underline{\textit{Effect of disorder }:} It is known that spin orbit interaction leads to suppression of the deleterious effects of disorder induced pair breaking on superconductivity\cite{mf}. In particular scalar disorder cannot mix states with different chirality. Stated differently scattering between different spin-momentum locked states acquire angular dependences arising from mismatch in spin orientation. The nontrivial dependence leads to vanishing dephasing rate yielding robust superconductivity\cite{mf}. 

\noindent\underline{\textit{Discussion }:} In this section we compare and contrast our work to those in the literature. To understand why only odd pairing superconductivity is obtained, it is important to note that the bands that touch are spin non degenerate. In other words, in the low energy effective theory there are two state per momentum which are split in energy. Chirality is a good quantum number but not spin. Given this, it is not possible to form spin singlets among degenerate states, as only one of the two "spin" states is available. Previous studies on the interplay of spin orbit and superconductivity \cite{gyc, meng} perform the mean field decomposition before projecting to the chiral basis. In other words a projection to singlet states is made before accounting for the splitting due to spin orbit. This allows for finite pairing amplitude among states that are non degenerate in energy in the noninteracting limit (i.e. mixes the valence and conduction bands). For chemical potential at the node these yield a class of even parity superconducting states for the $C_{4h}$ symmetric models. They are absent in the Weyl semimetals studied here.

Another important distinction is that in the minimal model assumed here of two Weyl nodes, the Pauli matrices represent spin. In particular they do not change under inversion. In certain class of effective theories, inversion operator takes the form $I:\sigma^{z}H(-k)\sigma^{z}$\cite{bb,gyc}. In this case the sign of the spin operators for the transverse directions changes under inversion. This additional symmetry leads to a set of superconducting states that allow for even parity spin singlet pairing. The reason is that the spin state at $\vec{k}$ and $-\vec{k}$ are no longer the same, as one would expect if inversion was an identity operator on spins. Thus there is a finite projection of singlet states onto the spin texture in the chiral basis.

A final point to note is that a full lattice model (as opposed to the low energy effective theory considered here) has linear dispersion for a finite energy window around the node. Thus any analysis that uses the full energy dispersion includes of the deviation from linearity. This is especially true for doped systems with large chemical potentials. Nevertheless the nondegeneracy of the bands and the spin structure allow for odd parity superconductors. Whether the even or odd parity states win out in this case is deferred to future investigations.

In summary Weyl semimetals are shown to display robust odd parity superconductivity, with both zero and finite momentum cooper pairs. We thank Chandra Varma for helpful discussion and comments.

\end{document}